# Sling-on-a-Ring: Structure for an elevator to LEO

Andrew Meulenberg and Timothy Poston

*National Advanced IPv6 Centre*
*Universiti Sains Malaysia*
*11800 Penang, Malaysia*
+601 25592369, US: 781-577-1594; *mules333@gmail.com*

**Abstract.** Various proposed space elevators may bypass the financial and environmental limits on rocket technology, but all have their own problems. A Low-Earth-Orbit (LEO) rotovator-based space-elevator version called "sling-on-a-ring" may overcome them. This mass-lifting system uses the spatial stability of an orbital ring, accessorized for transfer and storage of momentum and electrical power. A high-tensile-strength equatorial circum-terra loop of colossal-carbon tube (CCT) fiber has solar-power and station-keeping units and rotating sling modules. Long sling assemblies (~600 km) periodically descend into the atmosphere (to ~13 km). At perigee, the sling's rotational tip velocity almost cancels the orbital ring's velocity relative to Earth's surface. Split-second timing detaches a ~10-ton payload from an ordinary aircraft and jerks it into space by sling momentum, with the proven specific strength of CCTs now under development. This system eliminates the immense mass in space of other space-elevator systems, but needs extremely-long (100km) compressive members. Conceptual analysis for mass reduction of these structures is the subject of this paper.

## INTRODUCTION

The space-elevator system here is a key developmental stage of a multi-purpose, low-earth-orbit-based, system of rings around Earth (Meulenberg and Karthik Balaji, 2011): the "LEO ARCHIPELAGO", an environmentally-sound, economically-responsible, and technically-feasible solution for some of man's problems on earth and a foundation and stepping-stone for the move into space. The initial ring will be a broadband, fiber-optic, communication system (Meulenberg, Suresh and Ramanathan, 2008). It will also be a test bed for experience with the dynamics and control of large ring systems. Near-term financial return on investment is expected, growing with population and broadband demand.

The second ring uses ring-system stability to support a sling-based mass-transport system (Meulenberg, *et al.,* 2008), vital for any large-scale system in space. This solar-powered Sling-on-a-Ring will lift payloads from conventional aircraft into space, without needing the inefficient, expensive, and environmentally-harmful rocket-launch systems now moving mass into orbit. It gives stability, a solar power base, and massive energy storage at no extra financial or mass-in-orbit cost. With [electrodynamic (ED) tethers](), energy and momentum stored by speeding the ring can be retrieved directly, in lifting mass from the earth, or indirectly (via the ED tethers) as electrical energy. Energy from solar power, stored when the converters are exposed to sunlight, can be retrieved at any time and from any location on the rings without long conductive cables.

An effective [mass-lifter]() in place could enable an extremely-large-area 'shade-ring'. During initial construction, this multi-purpose shade-ring will reduce/remove the space debris from LEO. Extended into a higher-altitude or 'slant' orbit, it eliminates the radiation trapped in the Van Allen belts. By intercepting the belts above their 'mirror' points near the magnetic poles, slant-orbit shade-rings absorb, scatter, and slow the higher-energy trapped electrons and protons and nearly all lower-altitude radiation. With only quickly-suppressed transient radiation from solar flares, now-unviable near-earth orbits (*i.e*., MEO) become usable and even habitable. These shade rings can grow decade by decade until, by the end of the century, they reduce critical portions of the earth's temperature by several degrees Celsius (Meulenberg, Suresh and Ramanathan, 2009), from tilt orbits that shade portions of the near-polar sea ice and tundra from up to 10% of the summer sunlight. Preventing, or slowing, seasonal sea ice melting keeps the regional albedo high, and increases the impact of local shading by an order of magnitude. Keeping the tundra from melting prevents the mass frozen methane hydrates trapped there from escaping and competing with $CO_2$ as Earth's primary greenhouse gas.

Even as power needs in space increase, mass-lifting capabilities may exceed priority demand. As space experience with new solar-conversion technologies grows, specialized 'Power Rings' will occupy sun-facing near-dusk/dawn orbits (Meulenberg *et al.*, 2009). With 100% exposure to sunlight and a uniform environment, high-efficiency systems can provide laser power to mobile units operating within the LEO system and deep into near Earth space. As power-ring capability grows beyond space needs, mankind can choose either MEO-ring-based or Geostationary-based (or both) space-power systems to feed its insatiable need for power. The feasibility and cost effectiveness of this power ring, and of the shade ring, may depend strongly on a successful mass lifter.

Earlier papers addressed the enabling technology of tensioning capacity from the high specific strength of colossal carbon tubes for the sling-on-a-ring. This paper addresses the compressive members (spars) required to spread the ring into a balanced pair of lines through which a sling cycles safely (even while picking up a payload).

## "SLING-ON-A-RING"

The Sling-on-a-Ring system uses high-speed, globe-circling rings, at near free fall (zero-gravity), to store energy and to stabilize a sling-hub subsystem. It integrates a thin, high-tensile-strength, equatorial, circum-terra fiber with associated power and propulsion stations, and rotating-sling modules. It transfers a payload from the earth to LEO by a variant of the HASTOL (Bogar *et al.*, 1999) system, but via conventional aircraft and a skyhook-on-a-sling. Advantages are very low mass relative to the rings in (Birch, 1982) or to any non-HASTOL space elevator like (Raitt and Edwards, 2004) and independence of 'air-breathing hypersonic' aircraft. It does rely heavily on sling materials with a high specific strength. The proof of concept now available for fiber materials of sufficient specific strength for the single-stage Sling-on-a Ring moves this space-elevator concept beyond the "what-if" stage.

Basic to the sling dynamics is matching sling-tip velocity to the atmosphere. The clockwise sequence in Figure 1 shows the tip near Earth and the upper end of the tether, fixed on and moving with the sling 'hub' center of mass at about 7500*m/s* relative to the surface. At the point of pickup (13-15*km* above Earth), the tether-tip tangential velocity equals in size the sling's center of mass (COM) velocity (rightward in the figure). The net result (at sling-tip perigee) is near-zero tangential velocity of the sling tip relative to the transfer aircraft or to Earth. There is no fiery entry into the atmosphere. The actual rotational velocity of the sling can optimize: the horizontal tip velocity and stability relative to the payload-delivery aircraft (from -100 to +1000*km/hr* airspeed); the pickup window (altitude and timing); and payload acceleration after pickup.

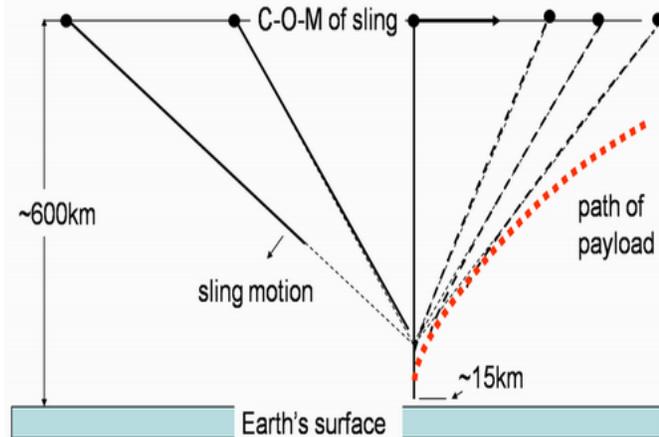

**Figure 1**. Sling dynamics.

The sling tip drops vertically into the atmosphere, and lifts the payload nearly vertically (red dashed path) out of it. An aircraft grapple of the sling tip 'falling' from space is a speed-up of film retrieval from early photo-reconnaissance spacecraft. Non-computerized, prop-driven aircraft met and caught small packages parachuting into the lower atmosphere. Mounting the grapple on the aircraft avoids raising tether mass, or reducing payload by this mass. A more elaborate and secure grapple bypasses the weight constraints of a sling- or payload-mounted system.

The intercept window is < 10*s* (unless various measures are taken to extend it), with little maneuver time. The tether is flexible, but under high tension. In CCT fiber, it must be ~6*mm* in diameter at the tip (for a 10*ton* payload) and thus locally approximates a thin rigid rod. Even with computer control, the dynamics are challenging, but the tip falls vertically through a low-turbulence region (at ~12*km*) as a body with active control surfaces for aerodynamic corrections by a computer on the aircraft that moves them both. The grapple coupling on the payload needs only to intercept any point on the tether, not its tip or other target (which would shrink the window to less than a second).

With the high vertical velocity, the aerodynamic body can move the tip laterally. This drag (or boost) can bow the sling as it approaches the payload, increasing the available pickup time and also buffering the payload-acceleration profile. At the intercept point, the tether can still be descending (which may double the time window relative to a rising tether), but the capture mechanism allows the tether to move freely up or down for the few seconds engaged. During this brief interaction, a counter-'cable' automatically wraps the tether in its 'rise'. Any wrap slippage, as the tether accelerates the payload upward from the aircraft, ceases when it meets the tip or another 'stop in the line.

The sling is integrated into an H-shaped high-strength, low-specific-mass, orbital ring assembly (Figure 2), the platform for the full system. It consists of a yoke, the harness H, a hub, the counter-mass (ballast), the tether (a 600*km* sling), and an aerodynamic body at the tether tip. The triangles join the H to the circum-terra ring. Among the objects of this study are 'spars' at the base of the triangles holding the H apart. If they collapse, the sling cannot pass through the ring, and catastrophic failure would result from the high energy in the sling rotation. Solar-power and station-keeping units are housed elsewhere on the ring, for balance. A slight spin-up puts the ring under tension, thus storing energy and momentum and enabling stable deployment and operation.

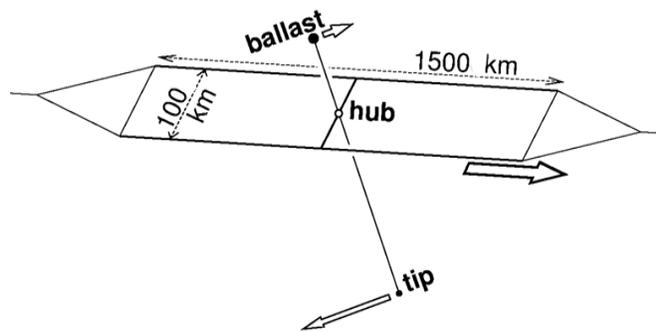

**Figure 2.** Stylized view of Sling-on-a-Ring.

The hub-to-ballast connection is semi-rigid. The ballast needs to move along part of this tether to balance angular momentum of the sling and payload. Rotation-induced tension rigidifies the hub-to-tip part of the flexible tether. A significant part (~50*km*) next to the hub is structurally rigid even without rotation (to assist tether spin-up and subsequent capture of the payload). After payload pickup, the sling C-O-M shifts toward the tip, and the angular velocity of the tip tether slows relative to the ballast tether — and more so at the tip. For relaunch of the tether for the next pickup, the high angular momentum and energy of the system and payload now transfer to the larger spinning ring (restoring some of the energy taken from it to lift the payload) via the harness and its connection with the hub, which rises slightly to accept the angular momentum (now about Earth). The rapidly shortening tether winds around until 'captured' at the ballast or hub (depending on which is chosen as its final destination).

After payload removal, the tip is ejected (by rail gun? – another rigid structure) to restart the sling. Applied stored energy and sufficient voltage make the sling and ballast 'self propel' as electrodynamic tethers.

The tensile strength and density of material of the sling are primary parameters. The specific strength (tensile strength divided by material density) determines the tether/payload mass ratio. (Meulenberg *et al.*, 2008) predicted that improving material strength would reduce tether mass enough to prove feasibility of the SOR within a decade.

## NEW MATERIALS

Most Sling-on-a-Ring structural components are in tension, with obvious benefit from new CCT-fiber technology. Only a few components must survive compressive forces and gain less obviously from the new material properties. These large and more-massive components are discussed in later sections.

Stronger materials yield a many-fold reduction in the tether/payload mass ratio. Based on technology expected in the near future, Meulenberg *et al.*, (2008) suggested a path to super-high-tensile-strength materials within a decade and to useful tether-payload mass ratios in space within 10 to 15 years from beginning of the project upon material validation. This required mass production of Carbon Nanotube (CNT) fibers with tensile strength of around 62*GPa* and density of 1.34*g/cc*. However, recent validation of a new high-specific-strength material, Colossal Carbon Tube, CCT, (Peng *et al.*, 2008), gives a new time line with CNT no longer needed. Useful tether-payload mass ratios could now be expected in space within 10 years from the project start.

CCTs are double-, or multiple-, layered thin sheets of graphene connected by mono-layered structures to give a thin, reinforced, hollow-core compound layer rolled into multi-micron-diameter tubes small enough to exclude from their interior any binder connecting them. Since the tube walls are nanoscopic, and the interior voids are microscopic, most of a CCT fiber's volume is nearly vacuum or low-pressure gas (as in an aerogel) from the fabrication process. Relative to the presently short, small-diameter tubes of CNT, CCTs are large, long, and nearly massless. Individual fiber densities have been calculated to be at 11*mg/cm$^3$*, with cell-wall densities of ~ 116*mg/cm$^3$* (Peng *et al.*, 2008).

In a composite, the large CCT surface areas and small relative-contact area need less binder than the nanoscopic structures of CNT, as it can 'wick' to the fiber contact points, leaving much inter-space void. Fibers from CCT have superior properties: extremely high specific strength (low density, as volume is mostly gas), excellent ductility (tube-dominated, rather than binder-dominated structure), and high electrical conductivity (from the graphene).

Even recent (2009) CNT macrofibers, assembled from individual nanotubes, show relatively-low tensile strength, in the low *GPa* range. CCT macrofibers, at <1/10 CNT the density, already have demonstrated a tensile strength of 6.9*GPa.* This confirms the much improved CCT mechanical properties in similar sized macrofibers of carbon nanotubes. A possible development time frame suggests mass-production quantities within the decade.

Owing to its architecture, an important feature of the CCT is its excellent ductility, very encouraging in situations requiring high toughness (as in the tether). The CCT deforms under tension in a ductile manner, with diameter shrinking in continuous local ways before breaking, much like the deformation of a metal wire or rubber band. This is in sharp contrast to the typical brittle-fracture nature of many advanced fibers. In fact, the CCT can sustain about 3% strain before failure, thereby giving adaptation to flaws in the fiber and warning before complete failure.

The combination of high strength and low density makes CCT a high-specific-strength, high-tenacity solution for the Sling-on-a-Ring (or any space-elevator). Its specific strength, about 15 times that of the strongest carbon fiber (T1000) and other capabilities now allow us to simplify the system architecture (Meulenberg *et al.*, 2009).

## CCT vs. CNT

Carbon nanotubes have been the latest dream material for high-tensile strength applications (like single-crystal iron fibers once were), but translation from microscopic to macroscopic has been a perpetual problem. However, no long fibers have been grown. The microscopic tensile strength of colossal carbon tubes is only about 1/9 that of CNT, but the very low density of CCT greatly raises its specific strength. The extreme theoretical tensile strength of CNT (~62*GPa*) is balanced in the relative tether-mass analysis by the low density of CCT (6.9*GPa*), giving similar results; but, with its higher specific strength, the CCT fiber is about twice as good as CNT (*i.e.*, tether-to-payload mass ratio is half that of the CNT for tip velocities >7000*m/s,* Meulenberg, *et al.*, 2009).

The high specific strength of both CCT and CNT fibers allows an increase in the tether's safety factor (sf). The major hazard to long tethers is micrometeoroids. In response, the tether sf can rise from 2 at the hub to 4 in the thin tip region and still have a useful T/P ratio. However, the CCT fiber, of low density and mainly empty space, is a natural shock absorber. Much of the micrometeoroid impact damage to normal materials is in the impact crater (several times larger than the micrometeoroid itself) and from spallation (a result of the pressure wave terminating abruptly on the far side of the material). For most materials, the type of micrometeoroid (metal or snowball) does not matter, but 'non-contacting' thin film layers are the best moderator of this concentrated energy source. Since many micrometeoroids shatter like snow balls on impact (almost independent of the impacted material), in CCT fibers their total energy will dissipate locally, but not microscopically, at the point of contact. The resulting damage mode must be explored. The resilience and low density of the CCT fibers should benefit this case by distributing the energy through a larger volume, but less destructively since less structural material is involved. It is believed that, at the other extreme, the damage in CCT macrofibers from the metal-ball type may be limited to a puncture hole not much larger than the incident micrometeoroid diameter. Both damage modes must be confirmed.

Table 1 has tether/payload ratios (critical to the timeline for sling operational deployment) of a 600*km* system for three high-tensile-strength materials, assuming ring-relative sling tip velocity 7000*m/s* (aircraft velocity bridges the gap to 7500*m/s*), a 10*ton* payload, and a graded sf. The CNT values are unvalidated 'hoped for' fiber values and theoretical limits. The low CCT density more-than-balances CNT's high tensile strength. A tether mass of 3 times payload, with CCT material validated in macrofibers and down from tether/payload of 100 for other materials, sharply changes perspective (notice radius at hub column).

**Table 1**. Tether-to-Payload Mass ratios.

| Material | Tether/Payload Mass | Radius at tip (mm) | Radius at hub (mm) |
|---|---|---|---|
| CNT (σ=20GPa, sf=4) | 100/1 | 1.8 | 110 |
| CNT (σ=62GPa, sf=4) | 5/1 | 1 | 3.8 |
| CCT (σ=6.9GPa,sf=4) | 3/1 | 3.1 | 8.5 |

## CCT in Structures

The benefits of the high specific strength of CCT tensile macrofibers are thus clear. Now, for the Sling-on-a-Ring, with the tether mass problem solved, the structural mass of the ring and sling components is dominant. Since the ring is in nearly free fall, the structural strength for many applications is greatly reduced. A major requirement for strength is in man-rated pressurized habitats and work areas. The general use of cylindrical and spherical geometries naturally puts these structures into the tensile mode, fitting the high specific strength of CCT. In microgravity, the ability to 'wind' structures in orbit out of CCT fiber may out-compete fabricating parts on Earth, transport to orbit in small holds through a very-stressful launch phase, and assembly in space. This latter option is just not viable for large structures that must later maintain internal pressures, which translate into very-large forces.

One potential technical problem is the mass associated with compressive members required for the SOR. For each sling, at least two 100*km* 'spreaders' (at the base of each triangle in Figure 2) keep the H sides apart. The actual compressive loads are not great, but buckling must be wholly suppressed. For best results, the H crosspiece should also be semi-rigid and self-supporting, and withstand large lateral loads with only moderate give. While much of the stability comes from tension, and thus benefit greatly from CCT, there are compressive elements that will be massive simply because of their large dimensions (100*km*). The compressive requirements could perhaps be best treated by use of 'tensegrity' structures based on tension with compression localized to isolated elements, and provide both lateral and longitudinal stability with a minimal mass. The mass of the compressive isolated members shrinks with CCT replacing the glass- or carbon-fiber-reinforced plastic in the tubes often used. Based on density and other properties, it is anticipated that the mass in the compressive elements (without loss of strength) could be 3–10 times less than with fiberglass or aluminum. Since the compressive members of the Sling-on-a-Ring may be its most massive components, the use of CCT can contribute in this area as well. The next section looks at tensegrity structures in space and the issues of optimizing them for SOR.

## TENSEGRITY

It is not novel to suggest the use of tensegrities in space. Skelton, Williamson and Har (2002) address their stability, and Zeiders, Bradford and Cleve (2004) suggest them as support for space mirrors and "a lightweight, deployable structure as an assembly of tensegrity modules like the one" in Figure 3. However, their use as rigid structures such as masts "is not very promising since these structures, pure tensegrity that is, are quite elastic and flexible; too much so for use as antennas with dishes mounted on top". From the main person, Snelson (1965), who has actually designed and built long tensegrity masts (most others follow his designs), these words are not encouraging. However, the reasons for this flexibility are not inevitable in tensegrities.

Figure 3 has 4 tensegrities. The space mast structure (a) in Zeiders, Bradford and Cleve (2004) merges copies of (b) from Snelson's original design. Figure 3c is Snelson's mast design. Note that the rods are not touching. Figure 3d is a highly-elastic form, easily made with rubber bands. The rods in (a) are not isolated; but unless joints are stiff, general tensegrity mathematics holds.

All the designs in Figure 3 (and most designs in the literature) have 'soft modes' when made with barely extensible cables. That is, particular load patterns can deform them strongly, with forces much less than their cable tensions. Only high pre-stress stiffens them against such loads. The structure in Figure 3b, for example, strongly resists force pushing together the rod ends on the left of the image, but the top twists easily relative to the bottom. Vertical compression produces mainly-twist yield. Similar behavior holds for any tensegrity with *N* rod ends joined by fewer

*Space, Propulsion & Energy Sciences International Forum (SPESIF-11, March 15-17, 2011)*

than $3N$-5 rods and cables: the examples in Figure 3 have $N$ = 9, 3, 18, 60 with 21, 9, 48, 120 rods and cables respectively. All Snelson designs studied here can thus be seen, simply by counting, to have soft modes.

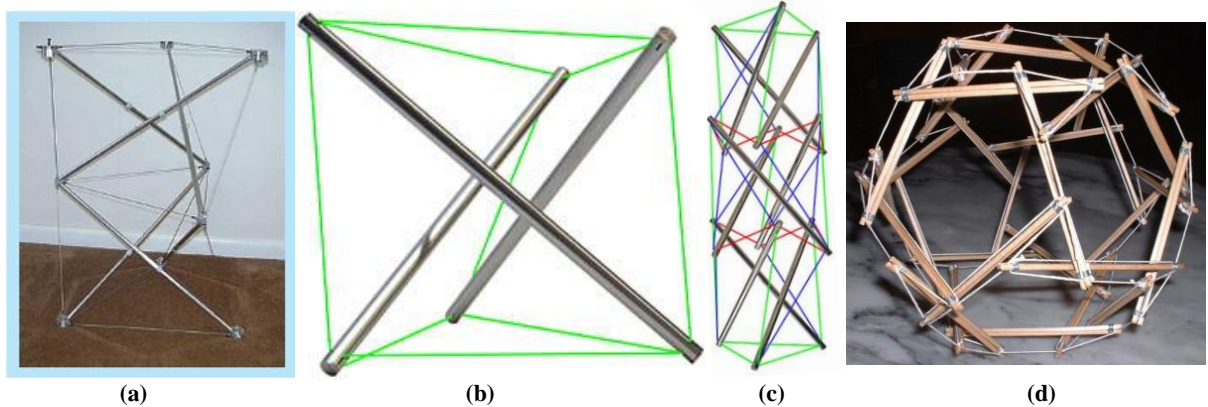

(a)   (b)   (c)   (d)

**Figure 3**. Space mast structures (a) Zeiders, Bradford and Cleve (2004) and (c) Snelson. (b) Snelson's building block. (d) A highly-elastic form (by permission Adrian Rossiter, http://www.antiprism.com).

It is useful to distinguish 'soft cables' like the rubber bands in Figure 3d, which are often held at several times their relaxed length, from 'tough cables', barely extended even at great tension. A tensegrity with soft cables does not have distinct soft modes, but deforms easily in many ways. When made with freely-chosen lengths, it settles to an equilibrium like Figure 3b, often arriving at a configuration that with tough cables has soft modes. This makes soft cable an easy way to find equilibrium geometries by experiment, but a bad way to find firm structures.

The energy minimizing of soft cables often leads to minima of particular lengths, subject to fixing the rest: but to first order a minimal length has zero change when the rest move, leading to a soft mode. (Burkhardt, 2008) makes such length minimization a design principle, with the inevitable consequence of soft modes and difficulty of building and maintenance: as he says of 3b, without careful length control "the structure will turn out to be a loose jumble of sticks and fishing line." In contrast, Figure 4a shows a 10-cable structure at different angles to the 9-string in Figure 3b. It forgives small length errors and, by tightening one cable (by pulling the free end through a rod) all are tightened. Furthermore, the structure has no soft mode and, thus, is much firmer. This is a general feature of our proposed class of designs, shared by Figure 4b that strongly resists deformation if made with low-extensibility cables. No force pattern on the rod ends allows a deformation where the cables twist away from the stress, more than they lengthen.

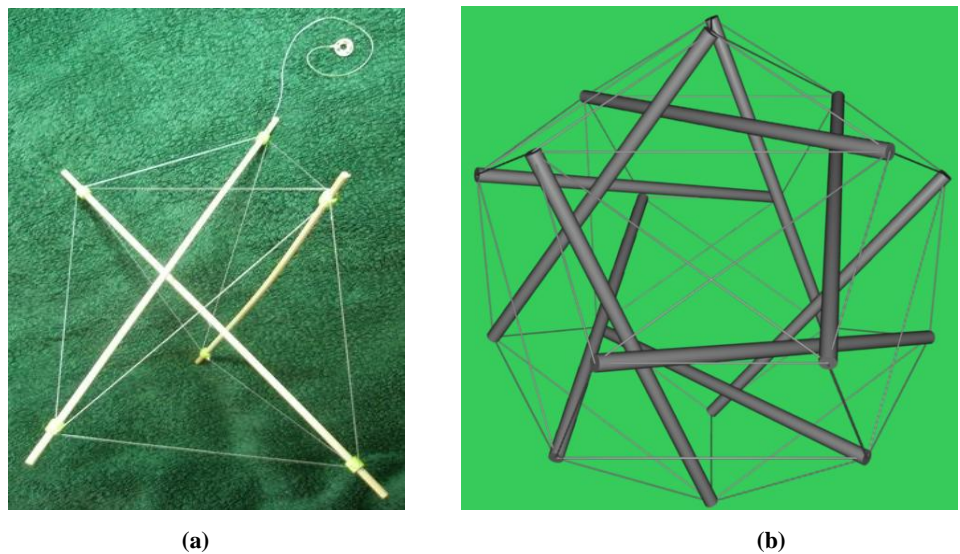

(a)   (b)

**Figure 4.** (a) A firm 10-string three-rod tensegrity, and (b) A 45-string dodecahedral design that strongly resists deformation.

The necessary "3*N*-5 joins" condition follows from a sufficient condition: when linearized, the constraints given by the rods and cables span the motion vectors possible for the rod ends with no net turn or translation, with **0** as a convex combination of the cable constraints modulo the rest. If this fails, the structure may still be stable, as in Figure 3, but second order terms decide it: the linear expansion is not enough. Any null direction of that expansion leads to a soft mode in the tough-cable limit, and resisting deformation in that mode requires raising the pre-stress tension in the cables. As they must support this tension even with no external load, they must be stronger and more massive. On Earth this leads to a gravitational load: resisting it needs more pre-stress, and thus heavier cables, in a vicious circle. In space, mass is a major cost factor in itself.

Minimizing length as in (Burkhardt, 2008), or cable count as in much of the literature, is thus a counter-productive design criterion. Less harmful is the "triangulated" goal in Snelson's work "it's possible to build such a structure whose network is non-triangulated. Such structures are flaccid and decidedly not firm." Nevertheless, the 12 points, 30 joins of "Tensegrity Tetrahedron; Total Triangulation" fail the "3*N*-5 joins" criterion and imply a soft mode. The rigidity of triangles, familiar from framework theory, is of less help where flaccidity is a non-local feature of the whole design.

A criterion is not a recipe, and the long structures that spanning convexity permits are still being explored. It is unlikely that a uniform mast does the best job of separating the H sides. Holding ends apart with a "strongest column" of solid material has a long and contentious literature, which maximizes load $\Lambda_c$ at first buckling. The best $\Lambda_c$ may not best meet the goal of keeping the ends *sufficiently* apart even when the separator is bowed but unbroken. The elastic resilience of CCT is important here, enabling structural integrity even for substantial deformations. A tensegrity's primary stiffness response involves cable stretching and some rod compression: for larger loads, individual rods may buckle. Slacking in a few cables — even breakage — need not be disastrous, as the load itself often stabilizes a tensegrity, reallocating tensions. Moreover, the goal need not be static separation of the H sides. The tether comes between them at highly predictable times, oscillating in synch with its ~6 turns/hour, and parting the sides only sometimes, like the Symplegades. The elastic resilience of CCT, in extension and compression, enables tensegrities to exploit this option in ways that classical structures cannot. Tensegrity structures can even be designed to exploit soft (twisting) modes as vast low-dissipation variable-'stiffness' springs that store energy and angular momentum.

## CONCLUSIONS

This paper presented a brief overview of the Sling-on-a-Ring concept as a mass lifter into LEO. New materials have major impacts on system feasibility, design, and costs. While earlier papers developed the benefits of the high specific strength of possible materials (such as colossal carbon tubes) in tension elements, the present work explores application in compression and a specific structure. The exceptional properties of the new material and its derivatives may make tensegrities more apt for large-scale space structures. With new concepts for optimizing these structures, tensegrities can become natural for space development.

## NOMENCLATURE

Pa = Pascal = SI unit of pressure (1 *newton/m²*)  
$r$ = density (*g/cm³*)  
σ = tensile strength (GPa)  
σ/$r$ = specific strength ($10^9$ *newton/(kg/m)*)

## ACRONYMS

CCT – Colossal Carbon Tubes  
CNT – Carbon Nanotubes  
C-O-M – center-of-mass  
EDT – electrodynamic tether  
HASTOL – <u>H</u>ypersonic <u>A</u>irplane <u>S</u>pace <u>T</u>ether <u>O</u>rbital <u>L</u>aunch  
LEO – Low Earth Orbit  
MEO – Medium Earth Orbit  
SOR – Sling-on-a-Ring  
T/P – tether-to-payload mass ratio  
sf – safety factor

## ACKNOWLEDGEMENT

This work was supported in part by HiPi Consulting, New Market, MD, USA, by the Science for Humanity Trust, Bangalore, India, and by the Science for Humanity Trust, Inc, Tucker, GA, USA.

## REFERENCES

Birch, P., "Orbital Ring Systems and Jacob's Ladders – I," *J. British Interplanetary Soc.*, **35**, (1982), p. 475.

Bogar, T. J., Bangham, M. E., Forward, R. L. and Lewis, M. J., "Hypersonic Airplane Space Tether Orbital Launch (HASTOL) System: Interim Study Results," AIAA paper, *9th Int. Space Planes and Hypersonic Systems and Technologies Conf.*, Norfolk, VA, USA (1999).

Burkhardt, R. W., *A Practical Guide to Tensegrity Design*, angelfire.com, (2008).

Meulenberg, A. and Karthik Balaji, P.S., "The LEO Archipelago: A System of Earth-Rings for Communications, Mass-Transport to Space, Solar Power, and Control of Global Warming," *Acta Astronautica* (acc'd for pub.), (2011) arXiv:1009.4043v1.

Meulenberg, A., Suresh, R., Ramanathan, S. and Karthik Balaji P. S., "Solar power from LEO?" Proceedings of *International Conf. on Energy and Environment*, Chandigarh, India, **39**, (2009), pp. 423-431.

Meulenberg, A., Karthik Balaji P. S., Madhvarayan, V., and Ramanathan, S., "Leo-Based Space Elevator Development Using New Materials and Technologies," Proc. of the *60th International Astronautical Congress,* Daejeon, (2009).

Meulenberg, A., Karthik Balaji P. S., Suresh, R., and Ramanathan, S., "Sling-On-A-Ring: A Realizable Space Elevator to LEO?" Proc. of the 59th *International Astronautical Congress (IAC-2008)*, Glasgow, Scotland, (2008).

Meulenberg, A., Suresh, R., and Ramanathan, S., "LEO-Based Optical/Microwave Terrestrial Communications," Proceedings of the *59th International Astronautical Congress*, Glasgow, Scotland, (2008) arXiv:1009.5506v1.

Peng, H., Chen, D., Huang, J. Y., Chikkannanavar, S. B., Hanisch, J., Jain, M., Peterson, D. E., K Doorn, S., Lu, Zhu, Y., Y. T., and Jia, Q. X., "Strong and Ductile Colossal Carbon Tubes with Walls of Rectangular Macropores," *Phys. Rev. Lett.* **101**, (2008), p. 14155.

Raitt, D. and Edwards, B., "The Space Elevator: Economics and Applications," Proc. of the *55th Int. Astronautical Congress* - Vancouver, Canada (2004).

Skelton, R. E., Williamson, D. and Han, J-H., "Equilibrium Conditions of a Class I Tensegrity Structure," *Advances in the Astronautical Sciences* 112, Spaceflight Mechanics, AAS 02-177, (2002).

Snelson, K. D., "Discontinuous Compression Structures," Patent No. 3,169,611, February, (1965)

Suresh, R. and Meulenberg, A., "A LEO-Based Solar-Shade System to Mitigate Global Warming," Proceedings of the *60th International Astronautical Congres*s, Daejeon, South Korea, (2009).

Zeiders, G. W., Bradford, L. J. and Cleve, R. C., "Lightweight Deployable Mirrors With Tensegrity Supports," *NASA Tech* Briefs, (2004).